# Line Profile Studies of Coronal Active Regions in Fe XII λ195.12 Using Hinode/EIS

Maya Prabhakar🟢 and K. P. Raju🟢
Indian Institute of Astrophysics, Koramangala Bengaluru 560034, Karnataka, India; maya.prabhakar@iiap.res.in, kpr@iiap.res.in
*Received 2022 January 2; revised 2022 March 18; accepted 2022 March 20; published 2022 May 23*

## Abstract

Coronal active regions are studied using Hinode/EIS observations in the EUV line Fe XII λ195.12 by analyzing their line profiles from 2006 December to 2019 December. The period covers the last 2 yr of solar cycle 23 and solar cycle 24 fully. Active regions are the main source of magnetic field in the solar atmosphere, important in its heating and dynamics. Line profiles were obtained from various active regions spread across the Sun on a monthly basis from which we obtained the intensity, line width, Doppler velocity, and centroid and examined their variation during the solar cycle. The histograms of the Doppler velocity and centroid show that they behave in six different ways with respect to the position of rest wavelength. In addition, the shifts in the centroid were found to be more compared to the Doppler velocity. The variation of the line width with respect to the Doppler velocity or the centroid mostly follows a second-degree polynomial. A multicomponent line profile is simulated to explain the difference in the behavior of the Doppler velocity and the centroid with respect to the line width. We also find that the intensity and the line width of the different data sets show a global dependence on the solar cycle with a good correlation. The implications of the results for the coronal heating and dynamics are pointed out.

*Unified Astronomy Thesaurus concepts:* Solar atmosphere (1477); Solar corona (1483); Solar active regions (1974)

## 1. Introduction

The solar corona is dominated by a variety of structures, such as active regions, coronal loops, bright points, coronal holes, etc. The heating of the coronal plasma to millions of degrees, the acceleration of the solar wind, and the reasons behind the dynamic solar atmosphere are not fully explained. There is a need for the physical connection between the mass motions in the solar atmosphere and the coronal heating and possibly the solar wind to be probed in more detail, to have a better understanding of the coronal dynamics. Hence, the nature of mass transport mechanisms in the solar atmosphere plays a key role in addressing this connection.

Many attempts were made to address the above problems by studying (1) the flows into the upper solar atmosphere from the lower solar atmosphere, (2) the mass motions and their magnetohydrodynamics in the corona, and (3) the possible connection between the convection and the dynamics of atmospheric regions. The line profile analysis is one such tool that gives parameters like the intensity, line width, Doppler velocity, and the centroid, which provide information on physical conditions such as density, temperature, mass motions, etc. Several interesting results were obtained from such analyses.

First, many kinds of shifts were observed in different regions in different lines. Raju et al. (1993) reported excess blueshifts in the coronal line profiles. The emission lines observed in the quiet transition regions were reported to be redshifted (Peter & Judge 1999; Brekke et al. 2000). The existence of a highly dynamic corona was established in the late 1990s after the Solar and Heliospheric Observatory and Transition Region And Coronal Explorer observations (Brekke et al. 2000). Further, blueshifts were also reported by Peter & Judge (1999) in three coronal lines. All these observations point to the mass motions present in the coronal heights whose emission lines are Doppler shifted. The asymmetry caused in the spectral lines was correlated with the coronal heating. Smaller mass motions driven from chromospheric regions can contribute to the hot plasma in the corona as well. Nanoflares are the other known sources of hot plasma in the coronal heights. Chromospheric or type II spicules (Beckers 1968; De Pontieu et al. 2009), magnetic reconnection (Klimchuk 2006; Parnell & De Moortel 2012), and dissipation of the Alfvén waves and magnetoacoustic shocks on magnetic flux concentrations (De Pontieu et al. 2004, 2007) have been shown to contribute significantly to the coronal heating and/or solar wind acceleration.

In recent years, several authors have reported the blueshifts, and these are interpreted in terms of nascent solar wind flow. Tian et al. (2010) observed signatures of nascent solar wind in a polar coronal hole with some significant blueshifts. Similarly, nanoflare heating was linked with solar wind by Patsourakos & Klimchuk (2006) and type II spicules by De Pontieu et al. (2009).

Blueshifts were considered to be signatures of slow solar wind sources by Hara et al. (2008) and Brooks et al. (2011) as well. McIntosh et al. (2012), who conducted studies using a combination of Hinode/Extreme ultraviolet Imaging Spectrometer (EIS) emission-line spectra and image sequences from the Solar Dynamics Observatory/Atmospheric Imaging Assembly, also report the line profile asymmetry and complexities involved in the coronal emissions at temperatures below ≈1 MK. Episodic heating events that are rooted in the chromosphere can play a significant role in filling the Sun's upper atmosphere with hot plasma (De Pontieu et al. 2009; McIntosh et al. 2010). Tian et al. (2021) talk about pervasive upflows, which are considered to be linked with the solar wind and to contribute to the coronal hot plasma.

Further, the coronal lines are known to have nonthermal line widths along with the expected thermal line widths (Delone & Makarova 1969, 1975; Chae et al. 1998). Excess line widths of the order of 15–20 km s⁻¹ were reported by Skylab and Solar Maximum Mission in the coronal forbidden lines. They are explained as indications of wave propagation causing coronal heating (Hollweg 1978; Parker 1988). Variation of these line widths above the limb with height has been reported differently by different instruments. The correlation between the Doppler velocity and the line width suggests that the outflows are composed of multiple components. Correlation between the Doppler velocity and the line width is also seen in coronal green line profiles from active regions (Raju et al. 2011).

The coronal active regions are the source of many large-scale eruptions, including the coronal mass ejections and the solar flares. These regions are also found to be present around the solar prominences in many cases. Active regions are found to peak during the maxima of the sunspot cycle. These bright regions appear prominently in X-ray and ultraviolet images of the Sun. Large outflows at the boundaries of the active regions have been discovered from Hinode/EIS observations in recent years (Sakao et al. 2007; Hara et al. 2008; Doschek et al. 2008; Brooks et al. 2011). Studies of the fast upflows (≈100 km s⁻¹) in the coronal active regions have shown asymmetry in the hot spectral lines. Tripathi & Klimchuk (2013) state that type II spicules or any such mass motions that cause instant heating and chromospheric evaporation do not contribute dominant plasma any greater than 0.6 MK to cause coronal heating. The series of tests that were carried out by Klimchuk (2012) to examine the role of plasma from type II spicules in the coronal heating also showed that spicules do not provide high-temperature plasma to the corona but may provide cooler plasma that may get further heated in the coronal altitudes. Recent observations by Macneil et al. (2019), Brooks et al. (2021), Yardley et al. (2021), and Harra et al. (2021) also report flows in the active region boundaries.

So keeping in mind the asymmetry seen in the coronal regions dominated by the blueshifts, their possible connections with the solar coronal heating, and the acceleration with the solar wind, we aim to study the line width variation in various coronal active regions and its possible interdependency with other parameters during the solar cycle. We also report other results found in the course of study of the line profile analysis. The high spatial and spectral resolution of Hinode/EIS (Kosugi et al. 2007) enables us to study the solar corona in great detail. We have studied how the line profiles vary in different active regions for a period of 13 yr from 2006 December to 2019 December. This period covers the last 2 yr of solar cycle 23 and solar cycle 24 fully. The aim of the paper is to study the coronal active regions using the emission-line profiles. In our earlier work (Prabhakar et al. 2019) we found some correlation of the line width with the Doppler velocity and the centroid. This is from the ground-based eclipse observations that lasted for just 3 minutes and 37 s. Hinode also provides a great opportunity to study the long-term behavior of the corona. So we have used about 13 yr of the data to study the solar cycle variation of line profile parameters.

In the following section, we present the data and analysis. Further, we present the results in Section 3, where we talk about the behavior of the line profile parameters, their interrelations, and the correlations between the parameters and the solar cycle. A discussion is provided in Section 4. Lastly, we present our conclusions in Section 5.

## 2. Data and Analysis

The Hinode mission[1] is dedicated to studying the dynamic Sun and its impacts on the terrestrial climate. It provides sequential data unaffected by the day and night cycles for studying the progression of particular events. The EIS on board Hinode generates spectra in the wavelength bands centered at 195 and 270 Å having a bandwidth of 40 Å. EIS images the Sun in 1″ and 2″ slits to produce line profiles and uses 40″ and 266″ slits to produce monochromatic images. EIS slits can be used to raster-scan required regions to generate spectroheliograms, as well as in sit-and-stare mode to observe regions for repeated exposures to produce high-cadence images (Culhane et al. 2007).

In our study, we have considered the coronal active region data sets from Hinode/EIS in the EUV line Fe XII λ195.12. It is a strong emission line that ranges between 194.84 and 195.36 Å and is formed at a relatively low temperature of 1.25 MK. One data set is considered for every month starting from 2006 October, when the instrument was launched, until 2019 December. However, we found that for the months of 2006 October and November, the data quality is very poor with missing pixels.[2] We obtained good-quality data sets from 2006 December onward. There are no data sets available for 3 months from 2018 February to April, when the EIS instrument turned off unexpectedly; it was turned back on in 2018 May. We have mostly considered the first available raster scan data of active regions in every month. In case of nonavailability of raster scans, a few sit-and-stare data are considered. In total, there are 154 data sets, out of which 5 of them are sit-and-stare images. This includes 47 1″ raster scans, 102 2″ raster scans, 4 1″ sit-and-stare data, and one 2″ sit-and-stare data.

The data sets are analyzed using EIS routines (EIS software notes[3]). The prepping of the data was done using the eis_prep[4] routine on IDL. This routine removes the pedestal and dark currents, electron spikes and hot pixels, and calibrates the data generating level 1 FITS files. These level 1 fits data are further fitted with Gaussian curves using the eis_auto_fit routine, which automatically handles the wavelength correction as well. It should be noted that the line width values are also corrected for the instrumental profile. We have fitted the data with a single Gaussian for our study using the above-mentioned routine. (However, it can also be used to fit double-Gaussian curves.) This gives the intensity versus wavelength (line profile) for every point of the data set. The peak intensity, Doppler velocity, half width, and centroid are extracted from the line profiles. The data are composed of the active regions observed on both the limb and the disk. Figures 1(a), (b), (c), and (d) give an example of the intensity, line width, Doppler velocity, and centroid maps for a typical data set observed on 2007-10-01T00:19:13Z, respectively.

## 3. Results

The parameters obtained from Gaussian fitting have certain interesting features. First, we show a line profile simulation to explain the behavior of the Doppler velocity and the centroid with respect to the line width in Section 3.1. This is followed by focusing on features of the parameters by analyzing their histograms in Section 3.2. We found that the line width has significant correlation with the Doppler velocity and the centroid, which we describe in Section 3.3. Lastly, to get an overview of how these parameters behave during a solar cycle,

---

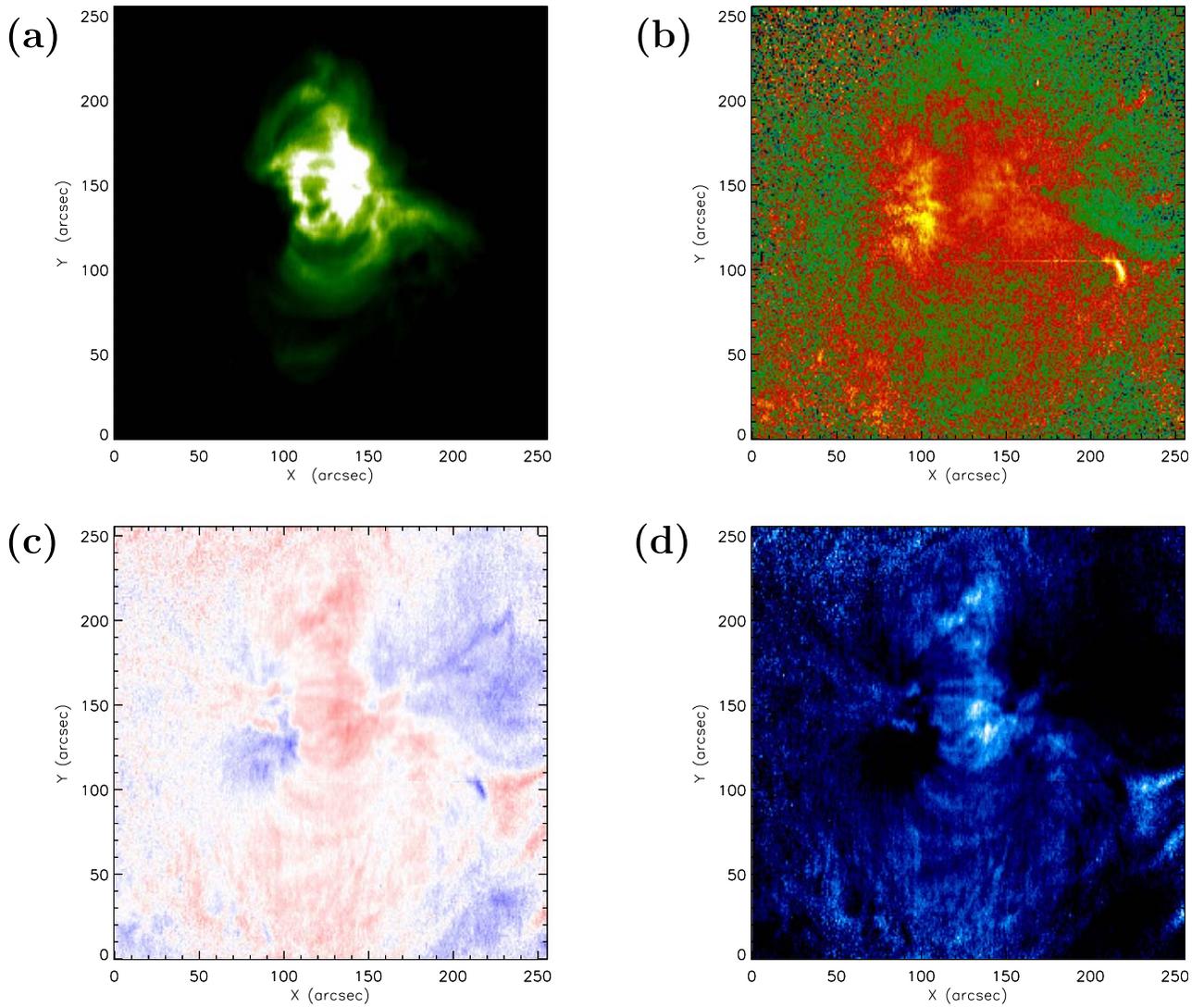

**Figure 1.** Maps of a typical 1″ raster-scanned active region's (a) intensity, (b) line width, (c) Doppler velocity, and (d) centroid observed on 2007-10-01T00:19:13Z having solar center coordinates [324″80, −73″43].

we address these parameters' correlation with the sunspot number in Section 3.4.

### 3.1. Simulations

Here we define Doppler velocity as the velocity corresponding to the wavelength of the peak intensity and the centroid velocity as the velocity corresponding to the wavelength point that divides the area of the line profile into two equal halves (Dadashi et al. 2011). When the line profile is a Gaussian, the two quantities are the same, but when there is asymmetry in the line profile, they tend to be different. This is shown through a simple simulation. A line profile is created by adding a stationary main component (intensity = 1 unit, velocity = 0 km s$^{-1}$, width = 35 mÅ) and a moving subsidiary component (intensity = 0.3 unit, velocity = 30 km s$^{-1}$, width = 35 mÅ). To make the profile look more realistic, we have added 2.5% random noise. The line profile is now fitted with a Gaussian, and then we obtained the Doppler velocity. The centroid velocity is calculated from the area of the line profile. The results are shown in Figure 2(a) and Table 1. It can be seen that the Doppler velocity and centroid velocity are different.

The calculations are repeated by changing the velocity of the secondary component in the range 0–60 km s$^{-1}$, and in each case the simulation was run 100 times. They show that the Doppler velocities and centroid velocities are not the same, especially when the velocities are large. Figure 2(b) shows the behavior of line width versus Doppler velocity/centroid. It can be seen that the variation is not linear but can be fitted with a higher-degree polynomial. As seen from our earlier paper (Prabhakar et al. 2019), a second-degree polynomial gives a good fitting, and there is no substantial improvement in a third-degree polynomial fit. It can also be seen that the estimated velocities and centroids in the single Gaussian fit are much smaller than the actual values. The estimated velocity decreases when the velocity of the secondary component is 60 km s$^{-1}$.

In the analysis, centroid is expressed in km s$^{-1}$ so that the comparison of the Doppler velocity and centroid is easier. Figure 2 gives an estimate of the error as well. The maximum error obtained from the simulation is 0.23 km s$^{-1}$ in the Doppler velocity and 0.76 km s$^{-1}$ in the centroid. However, a more realistic calculation considering the instrumental parameters gives the error as 1.6 km s$^{-1}$ (Dere et al. 2007).

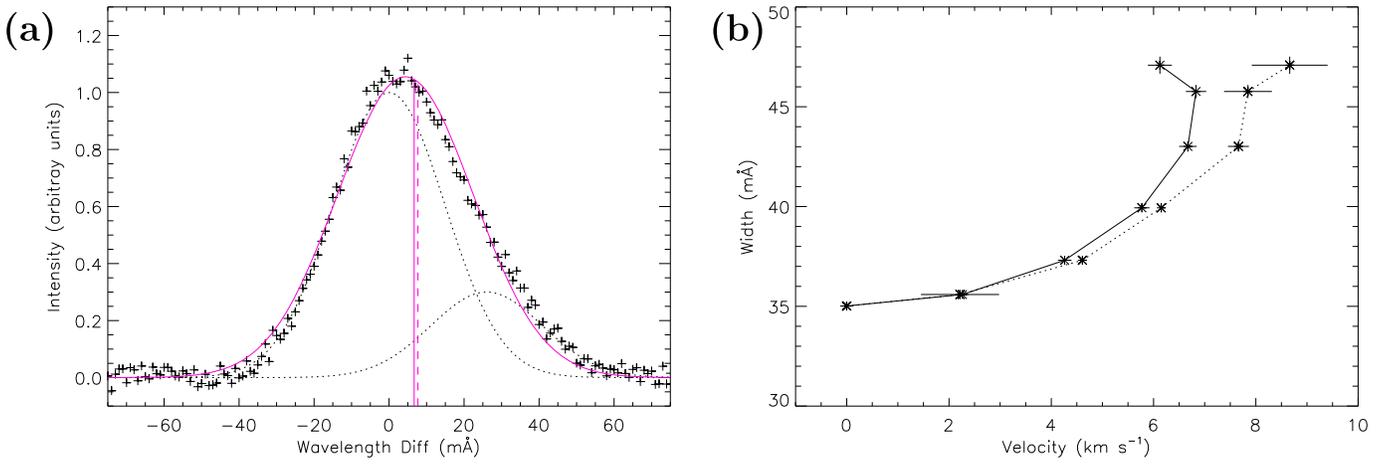

**Figure 2.** (a) Simulation showing a line profile with a stationary main component and a moving subsidiary component with 2.5% random noise fitted with a Gaussian. The vertical lines show the corresponding Doppler velocity (solid) and centroid (dashed). Details are given in Table 1. (b) Plot showing the variation of the line width with Doppler velocity (solid)/centroid (dotted).

**Table 1**
Values of Intensity (Arbitrary Units), Velocity (km s$^{-1}$), and Width (mÅ) for the Simulated Main and the Subsidiary Components with the Values of Velocity, Centroid (mÅ), and Width for the Fitted Curve

| No. | Component | | | | | | Fitted Gaussian | | |
|---|---|---|---|---|---|---|---|---|---|
| | Main | | | Sub | | | | | |
| | Int | Vel (km s$^{-1}$) | Wid (mÅ) | Int | Vel (km s$^{-1}$) | Wid (mÅ) | Vel (km s$^{-1}$) | Cen (mÅ) | Wid (mÅ) |
| 1 | 1.0 | 0.0 | 35 | 0.3 | 0.0 | 35 | 0.002 ± 0.12 | 0.0 ± 0.00 | 35.01 ± 0.18 |
| 2 | 1.0 | 0.0 | 35 | 0.3 | 10.0 | 35 | 2.26 ± 0.11 | 2.21 ± 0.76 | 35.59 ± 0.19 |
| 3 | 1.0 | 0.0 | 35 | 0.3 | 20.0 | 35 | 4.26 ± 0.13 | 4.60 ± 0.06 | 37.31 ± 0.21 |
| 4 | 1.0 | 0.0 | 35 | 0.3 | 30.0 | 35 | 5.77 ± 0.14 | 6.15 ± 0.00 | 39.93 ± 0.22 |
| 5 | 1.0 | 0.0 | 35 | 0.3 | 40.0 | 35 | 6.67 ± 0.17 | 7.65 ± 0.20 | 43.10 ± 0.26 |
| 6 | 1.0 | 0.0 | 35 | 0.3 | 50.0 | 35 | 6.83 ± 0.19 | 7.84 ± 0.46 | 45.76 ± 0.32 |
| 7 | 1.0 | 0.0 | 35 | 0.3 | 60.0 | 35 | 6.12 ± 0.23 | 8.66 ± 0.74 | 47.08 ± 0.42 |

### 3.2. Histogram Analysis

We report the behavior of histograms of the intensity, Doppler velocity, line width, and centroid of all the data sets in this section. Based on the shift in the Doppler velocity and the centroid, we have classified the data into six different cases. For each case, we have provided a typical example of the histograms. The intensity and the line width histograms are given for the sake of completeness.

*Case 1—Doppler velocity and centroid centered at the rest wavelength.* In this case, Doppler velocity and centroid are both centered at the rest wavelength. When the shifts are found within the error limits, such data sets are put under this category. The intensity and the line width histograms are skewed toward the higher values. We have observed 38 data sets under this category, which composes 24.6% of all the data sets analyzed. The one observed on 2007-04-01T06:12:21Z is shown in Figure 3.

*Case 2—Doppler velocity and centroid both centered in the blue wing.* Here both the Doppler velocity and the centroid histograms are shifted to the blue wing (Figure 4). The Doppler velocity ranges between −32 and 0 km s$^{-1}$ and peaks around −14 km s$^{-1}$. The centroid ranges between −30 and 0 km s$^{-1}$ and peaks around −13 km s$^{-1}$. The intensity distribution is highly skewed, whereas the width distribution is almost symmetric. We have observed only one data set under this category that is observed on 2017-04-02T02:15:13Z.

*Case 3—Doppler velocity and centroid both shifted to the red wing.* This is complementary to Case 2, where both the Doppler velocity and the centroid histograms are shifted to the red wing. We have observed only two data sets under this category; the one observed on 2008-06-01T09:50:26Z is shown in Figure 5. The Doppler velocity ranges between −3 and 18 km s$^{-1}$ with a peak at 6 km s$^{-1}$. The centroid ranges between −4 and 17.5 km s$^{-1}$ with a peak at 5.5 km s$^{-1}$. The figure also shows an intensity histogram with multiple components and a line width histogram that is skewed toward lower values.

*Case 4—Doppler velocity centered at the rest wavelength and centroid shifted to the blue wing.* In this case, the Doppler velocity histogram is centered at the rest wavelength, while the centroid is shifted to blue. The example shown in Figure 6 shows the histograms for the data set observed on 2018-05-26T13:03:57Z, where the centroid is shifted to the blue wing fully. The Doppler velocity ranges between −18 and 17 km s$^{-1}$ with a peak near the rest wavelength. However, the centroid ranges between −93 and 58 km s$^{-1}$ with a peak near −74 km s$^{-1}$, which is comparatively higher than all other cases, where it ranges between −30 and 0 km s$^{-1}$. This example shows an extremely blueshifted centroid. We see 71 data sets (46.1%) under this category. A majority of the data sets analyzed fall under this case. The intensity histogram is skewed toward higher values, and the line width histogram is almost symmetrical.

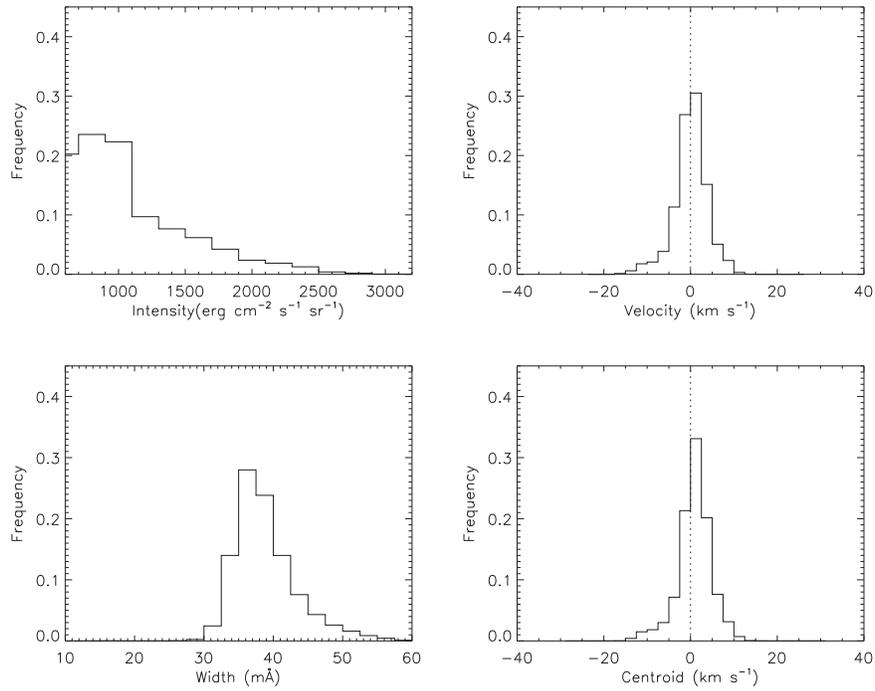

**Figure 3.** *Case 1.* Doppler velocity and centroid centered at rest wavelength. Top panel (left to right): histograms of intensity and Doppler velocity. Bottom panel (left to right): histograms of line width and centroid. The dotted line represents the zero shift.

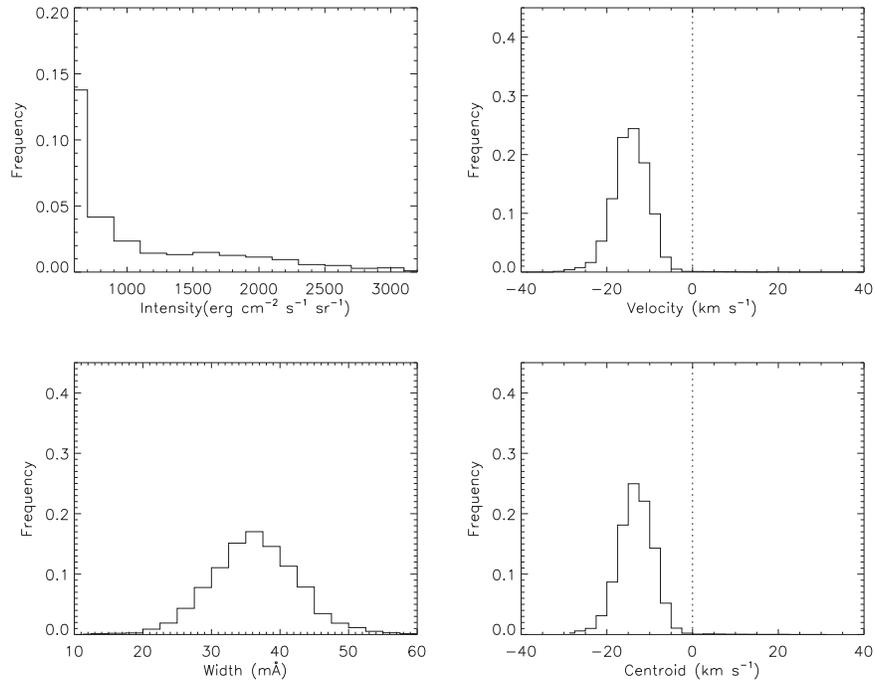

**Figure 4.** *Case 2.* Doppler velocity and centroid both shifted to blue completely. Histograms of intensity, Doppler velocity, line width, and centroid are as in Figure 3.

*Case 5—Doppler velocity centered at the rest wavelength and centroid shifted to the red wing.* Case 5 is complementary to Case 4, where the Doppler velocity is centered at the rest wavelength, while the centroid is shifted to the red, partially or completely. Figure 7 shows a typical example where the velocity histogram is peaked near rest wavelength and the centroid is shifted to red completely. The Doppler velocity ranges between −9.5 and 15.5 km s$^{-1}$. The centroid ranges between 12 and 37 km s$^{-1}$ and peaks around 22 km s$^{-1}$. This data set is observed on 2010-12-04T23:57:52Z. The intensity histogram is skewed as usual and has multiple components, and the line width histogram is almost symmetric. This case constitutes the second-largest category of data sets analyzed (26.62%).

*Case 6—Doppler velocity shifted to the red wing and centroid shifted to the blue wing.* This is the last category in the data sets where we see the shifts in the Doppler velocity and centroid in the opposite directions. The Doppler velocity is shifted to the red, while the centroid is shifted to the blue. We observe only one data set under this category on 2016-01-02T03:25:25Z, which is presented in Figure 8. Both the intensity and the line width are skewed here. The Doppler velocity ranges between $-25$ and $20\,\mathrm{km\,s^{-1}}$ and peaks at $3\,\mathrm{km\,s^{-1}}$. It is mostly in red, with a small tail in the blue wing.

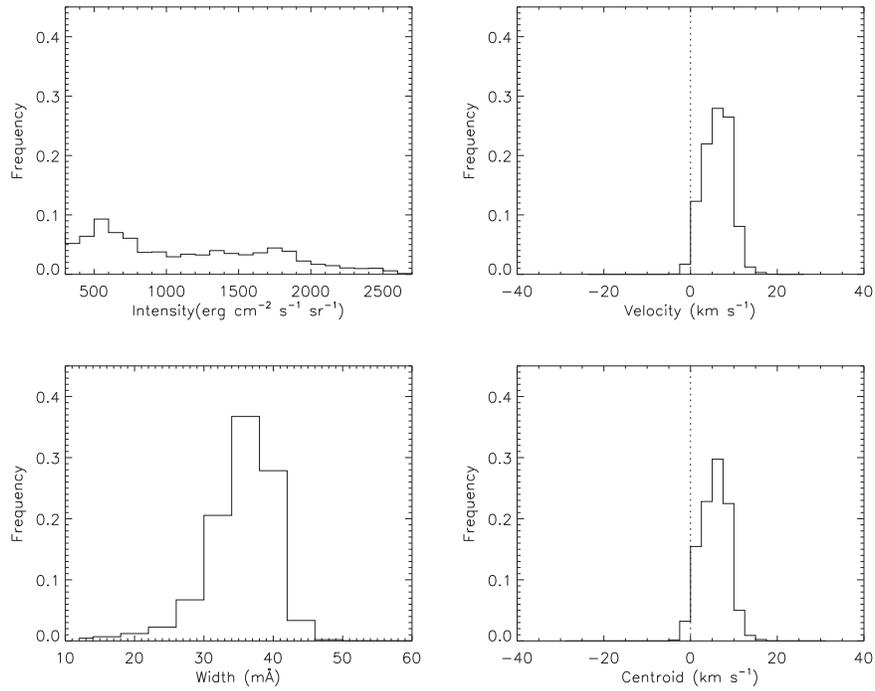

**Figure 5.** *Case 3.* Doppler velocity and centroid both shifted to red mostly. Histograms of intensity, Doppler velocity, line width, and centroid are as in Figure 3.

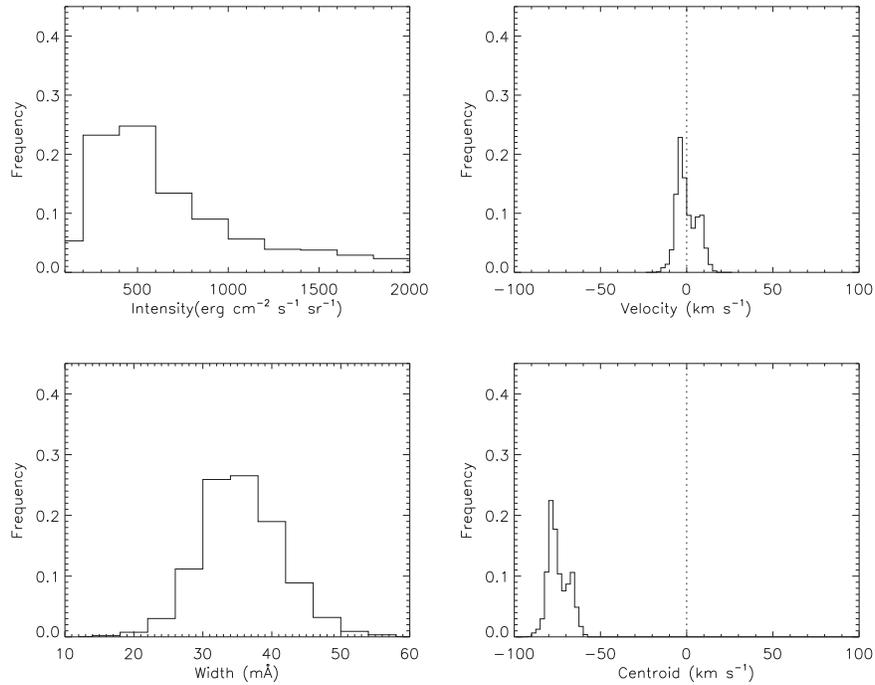

**Figure 6.** *Case 4.* Doppler velocity centered at the rest wavelength and centroid shifted to blue fully. Histograms of intensity, Doppler velocity, line width, and centroid as in Figure 3.

However, the centroid, which peaks at $-13\,\mathrm{km\,s^{-1}}$, is fully in the blue wing ranging between $-40$ and $0\,\mathrm{km\,s^{-1}}$.

As we have considered only one data set per month, we could have missed examples of different cases. There is also the possibility of a case complementary to Case 6 where the Doppler velocity is shifted to the blue and the centroid is shifted to the red. The results of the above analysis are summarized in Table 2.

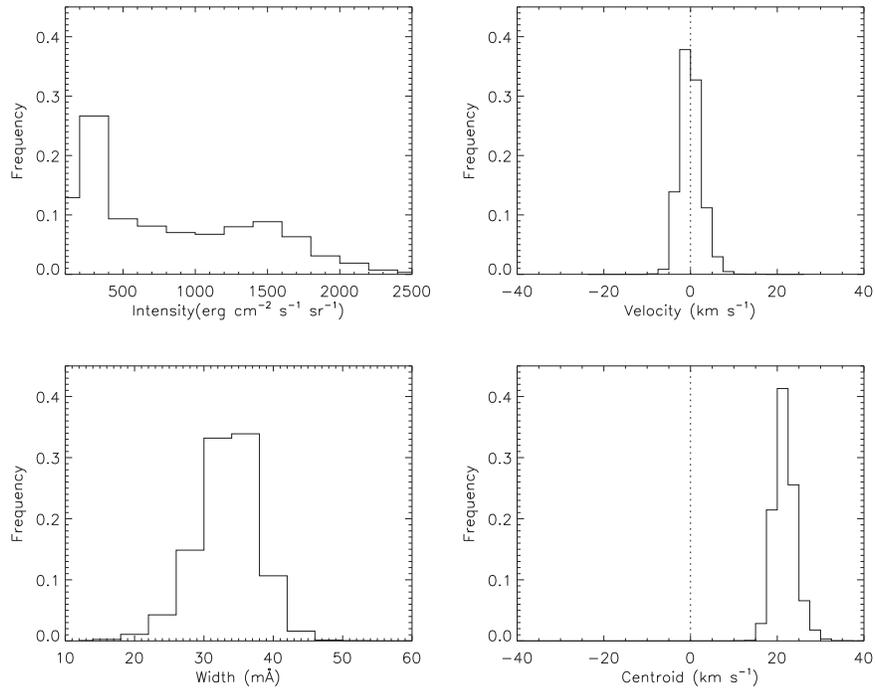

**Figure 7.** *Case 5.* Doppler velocity centered at the rest wavelength while the centroid is shifted to red completely. Histograms of intensity, Doppler velocity, line width, and centroid are as in Figure 3.

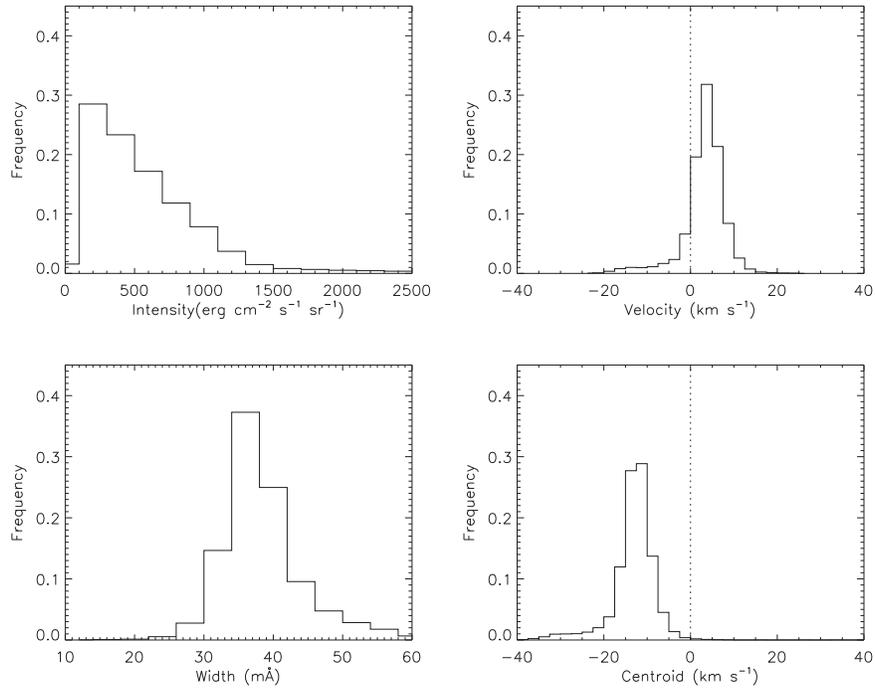

**Figure 8.** *Case 6.* Doppler velocity shifted to red largely, while the centroid is shifted to blue completely. Histograms of intensity, Doppler velocity, line width, and centroid are as in Figure 3.

### 3.2.1. Cases of Double Peaks

In this section, we report the special data sets where we have observed double peaks in the histograms of Doppler velocity and centroid. The data sets reported here fall under the cases mentioned above. However, their uniqueness comes from having two peaks, unlike what is seen in the previous section. The intensity and line width distributions behave as in the above cases. We report three examples here where we see such behavior.

Figure 9 taken from the data set observed on 2017-08-01T01:02:41Z is an example where Doppler velocity and centroid distribution have two peaks, one in the red and one in the blue wing. Hence, this can be considered as a special case of Case 1. It may be noted that no double peaks are observed in intensity or the line width.

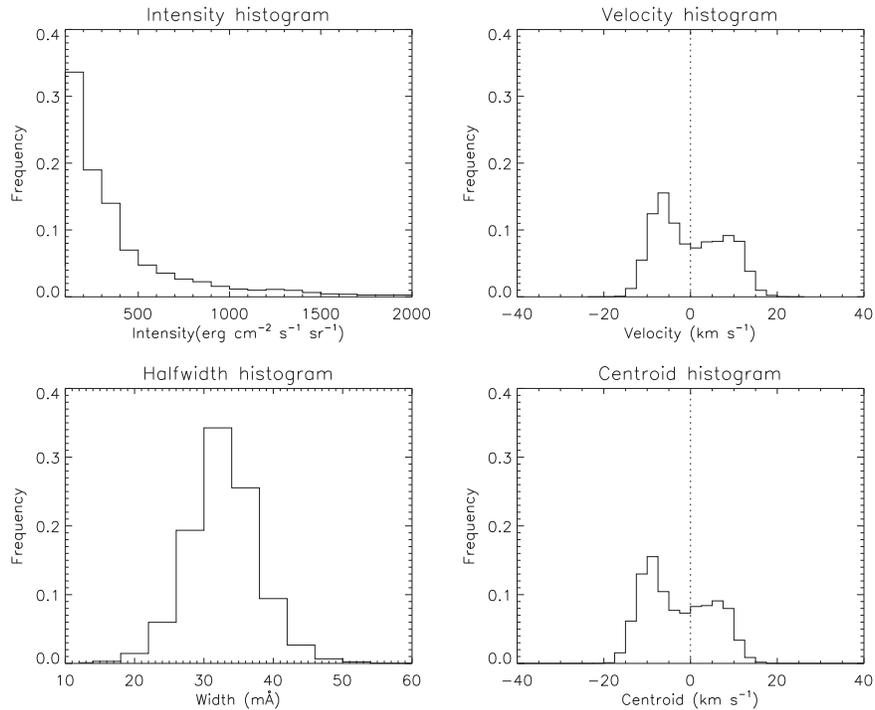

**Figure 9.** Double peaks seen in Doppler velocity and centroid, both centered at the rest wavelength. Histograms of intensity, Doppler velocity, line width, and centroid are as in Figure 3.

**Table 2**
Number of Data Sets Observed in Different Cases

| Case | Behavior | | No. of Data Sets | Percentage |
| --- | --- | --- | --- | --- |
| | Doppler Velocity | Centroid | | |
| 1 | No shift | No shift | 38 | 24.6 |
| 2 | Blueshifted | Blueshifted | 1 | 0.64 |
| 3 | Redshifted | Redshifted | 2 | 1.29 |
| 4 | No shift | Blueshifted | 71 | 46.1 |
| 5 | No shift | Redshifted | 41 | 26.62 |
| 6 | Redshifted | Blueshifted | 1 | 0.64 |

Figure 10 taken from the data set observed on 2018-06-01T03:29:43Z is a special case of Case 4. Single peaks are observed in intensity and line width histograms. The Doppler velocity and centroid histograms have two peaks. The Doppler velocity ranges between −23 and 21 km s$^{-1}$, with one peak in the blue wing and one in the red wing. The centroid ranges between −62 and 16.5 km s$^{-1}$ with an average of −38 km s$^{-1}$ having two peaks in blue.

Figure 11 is presented from the data set observed on 2019-07-04T00:05:13Z and is a special case of Case 5. The Doppler velocity ranges between −26 and 23 km s$^{-1}$ and has two equal peaks, one in the red wing and one in the blue wing. The centroid is shifted mostly to the red, ranging between −15 and 34 km s$^{-1}$. Again, there are no double peaks in intensity or line width.

### 3.3. Interdependence of the Parameters

Further, we examined the interdependence among the parameters. In particular, we studied how the Doppler velocity and the centroid of each data set vary with line width. Figures 12(a), (b), (c), and (d) give examples of line width versus Doppler velocity and line width versus centroid distributions of different data sets. We have earlier studied the correlation between these quantities from ground-based eclipse observation of the coronal green line Fe XIV λ5303 and found that a quadratic polynomial gives the best fit (Prabhakar et al. 2019). Therefore, we have tried the quadratic polynomial fits here. The details of the fits are shown in Table 3. The plot of Doppler velocity/centroid versus line width from simulations shows that the variation follows a polynomial-like behavior. In the first two cases, the trend is clear and the fitting is good, with the value of correlation coefficient $R$ being 0.72 and 0.69, respectively. This shows that the line width increases with the absolute value of Doppler velocity/centroid. In the third case also, similar trends are seen ($R = 0.67$), but there seem to be two different distributions involved. Figure 12(d) shows no correlation between the quantities. In most of the other cases, the trend is similar to those shown in Figures 12(a) and (b), but with a reduced correlation coefficient.

### 3.4. Correlation with the Solar Cycle

Here we examine the dependence of the parameters on the solar cycle. Note that the EIS observational windows are located at different spatial positions over the Sun. Our aim is to find any global dependence of the parameters on the solar cycle. So we correlated the intensity, line width, Doppler velocity, and the centroid with the monthly average sunspot number. The sunspot number is taken from Solar Influences Data analysis Center. Since there is a time difference between the two observations, the sunspot number is interpolated for the EIS observation times. Figure 13(a) shows the temporal variation of intensity and the sunspot number. The smoothed curve (red) is obtained from a five-point boxcar averaging. We can see that the variations of both the quantities are alike.

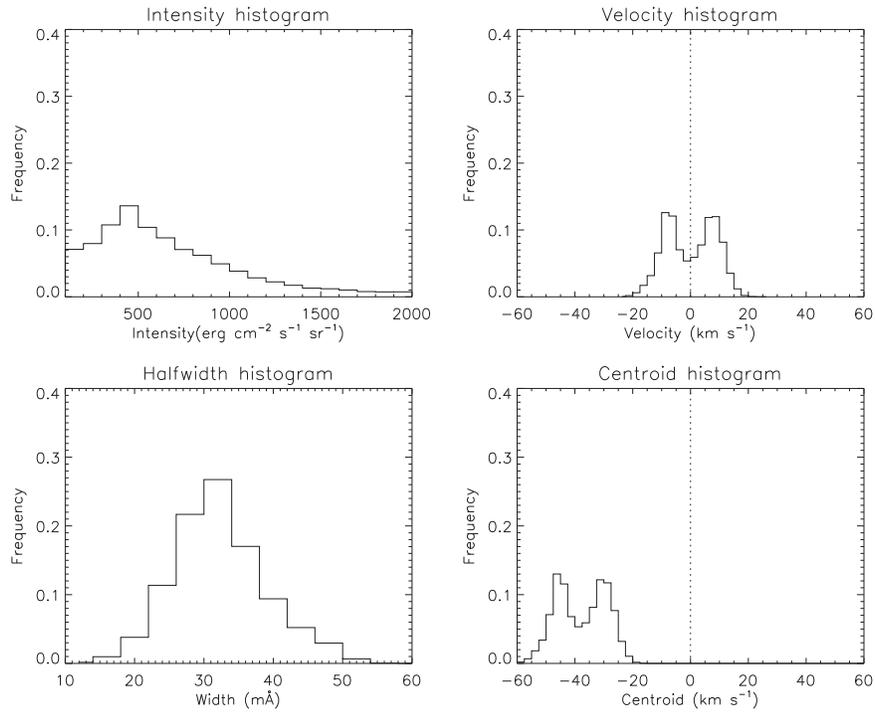

**Figure 10.** Doppler velocity is centered at the rest wavelength with double peaks. The centroid, also having double peaks is blueshifted. Histograms of intensity, Doppler velocity, line width, and centroid are as in Figure 3.

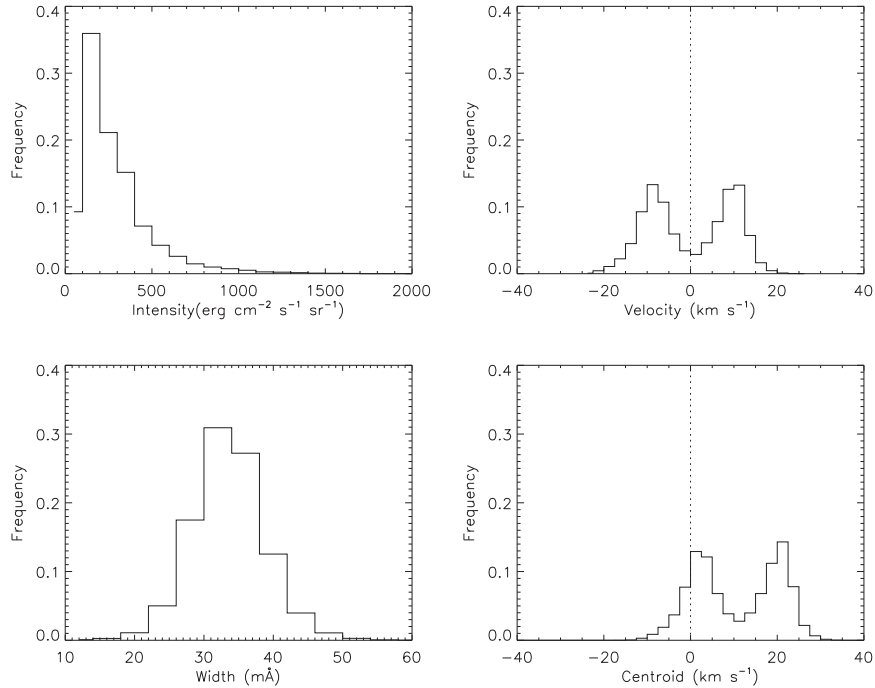

**Figure 11.** Doppler velocity is centered at the rest wavelength with double peaks showing no shift. The centroid, also having double peaks is largely redshifted. Histograms of intensity, Doppler velocity, line width, and centroid are as in Figure 3.

Figure 13(b) gives the cross-correlation between the two quantities as a function of lag given in years. The cross-correlation coefficient and the probability are given in Table 4. The maximum correlation coefficient ($R$) is 0.49, and the probability of the correlation occurring by chance is close to zero. The maximum correlation occurs at zero lag, and there is also an asymmetry with respect to zero lag that could be due to the difference in the rising and falling phases of the solar cycle. The solar cycle variation of intensity is obvious. Similarly, Figures 14(a) and (b) explain the same for the line width. Here the maximum correlation is −0.27, which occurs at a lag of around −3 yr, and the probability of the correlation is 0.001. It can be seen that the line width is weakly dependent on the solar cycle. The correlation is moderately significant because of the large number of observations (Taylor 1982), which may imply that there could be a dependence of line width on solar cycle with some lag.

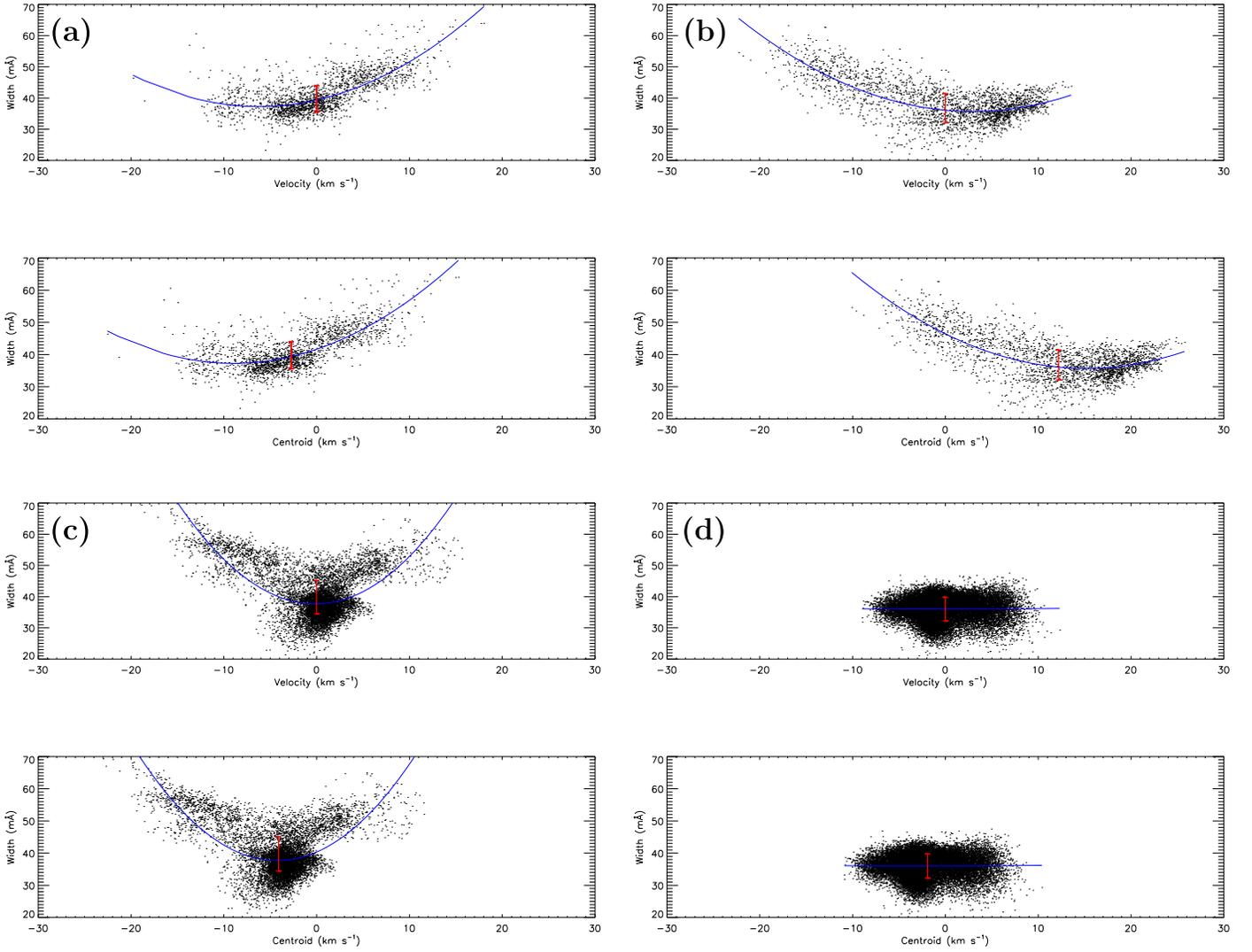

**Figure 12.** (a) Trend increasing in red having $R = 0.72$, (b) trend increasing in blue having $R = 0.69$, (c) symmetric trend having $R = 0.67$, and (d) constant trend having poor $R$. Top panel in each subfigure: variation of line width with Doppler velocity. Bottom panel: variation of line width with centroid.

**Table 3**
Quadratic Correlation Coefficients for the Examples Shown in Figure 12

| Figure | Observation (UTC) | | Quadratic $R$ |
|--------|------|------|------|
| | Date | Time | |
| 12(a) | 2012-07-05 | 18:18:27 | 0.72 |
| 12(b) | 2011-05-01 | 09:06:05 | 0.69 |
| 12(c) | 2011-11-01 | 22:58:03 | 0.67 |
| 12(d) | 2007-11-01 | 06:32:06 | 0.005 |

## 4. Discussion

Studies of the coronal spectra in recent years have revealed that there is a ubiquitous asymmetry present in the coronal spectral line profiles. Several works have reported redshifts and blueshifts in Doppler velocity, especially in the active regions. Chae et al. (1998) and Peter & Judge (1999) reported blueshifts in the coronal line profiles. Marsch et al. (2008) reported blueshifts in the active region boundaries in the EUV $\lambda195.12$ line and connected these blueshifts to the propagating disturbances. More often, the stronger blueshifts are reported in the hotter lines ($T \approx 1–2$ MK), which are found to increase as a function of temperature. Chen et al. (2010) reported blueshifts of the order of 20 km s$^{-1}$. From EIS observations, Tian et al. (2011) observed the main component of the line profiles to be stationary and the secondary component to be blueshifted. Raju et al. (2011) reported multiple components with an excess of blueshifts in the coronal green line profiles.

In Prabhakar et al. (2019), these authors have reported 59% blueshifts and only 7% redshifts of the Doppler velocity in the coronal green line profiles. However, in the present study, we see an almost equal distribution of blueshifts and redshifts in the Doppler velocity. We find that the shifts are mostly seen in the centroids. As indicated through the simulation, this could be due to the strong asymmetry in the long profiles. It may also be noted that the formation temperature of Fe XII $\lambda195.12$ is 1.25 MK, whereas that of the coronal green line is 2 MK. The results also imply that the Doppler velocity and centroid distributions are quite different.

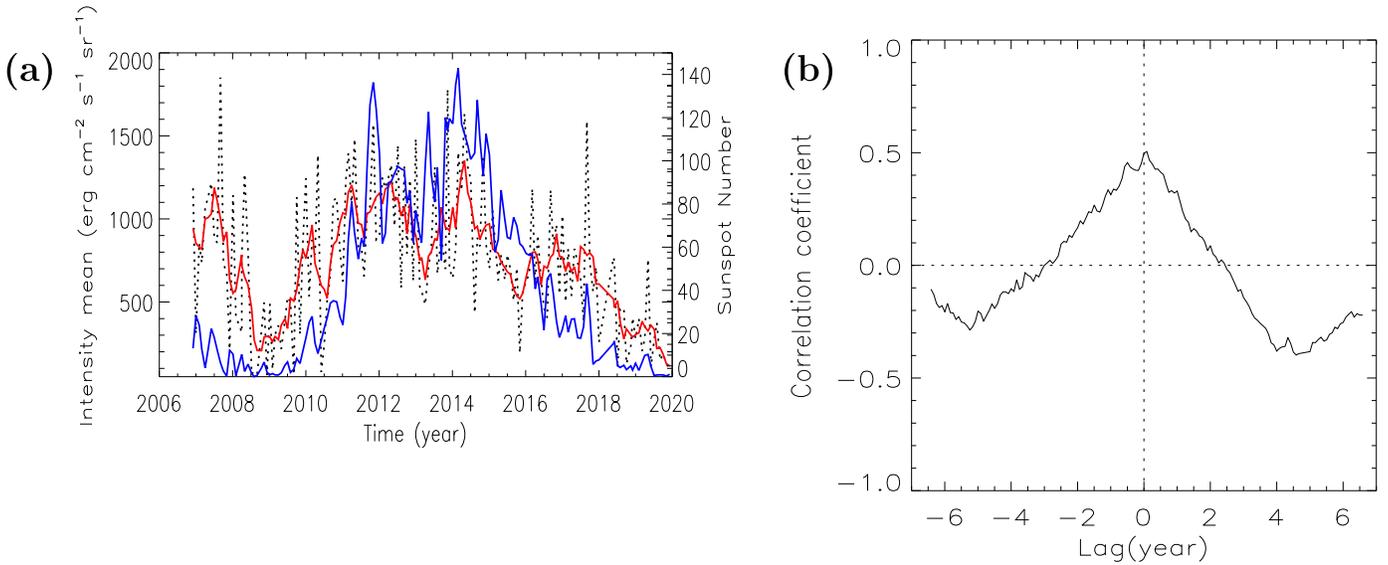

**Figure 13.** (a) Temporal variation of intensity (black) and sunspot number (blue), with the smoothed curve (red) showing the trend of temporal variation of intensity. (b) Cross-correlation of intensity with the sunspot number as a function of lag.

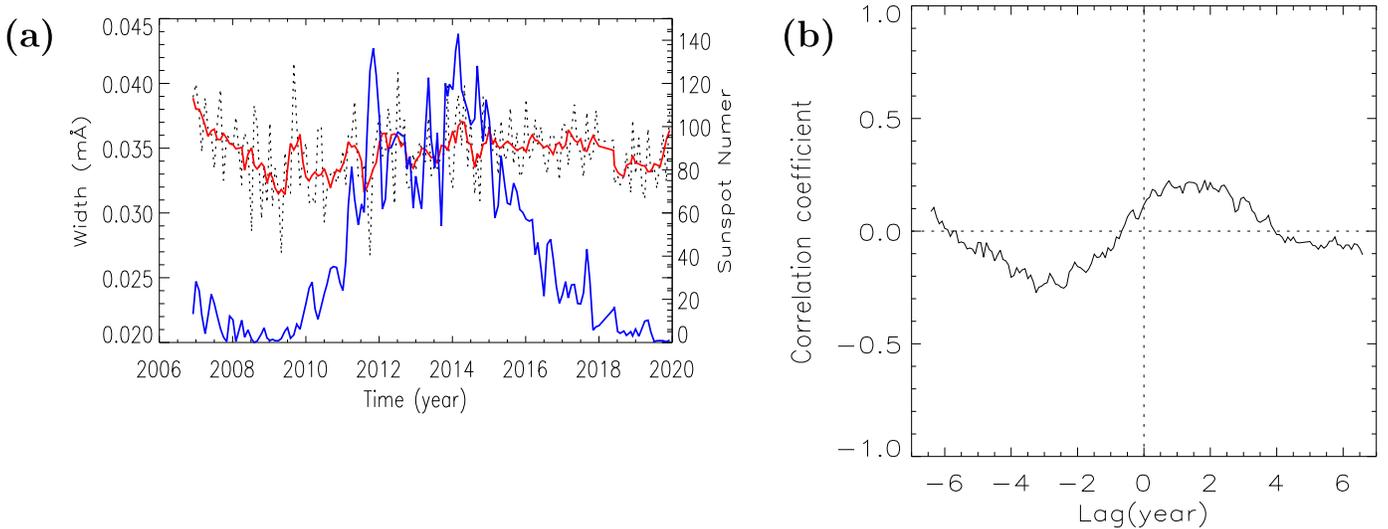

**Figure 14.** (a) Temporal variation of line width (black) and sunspot number (blue), with the smoothed curve (red) showing the trend of temporal variation of line width. (b) Cross-correlation of line width with the sunspot number as a function of lag.

**Table 4**
Maximum and Minimum Values of the Cross-correlation Coefficient of Variation of Intensity and Line Width with Sunspot Number with Time as a Function of Lag and the Probability of the Correlations

| Parameter | Figure | $R$ | | $p$ |
|---|---|---|---|---|
| | | $R$ max | $R$ min | |
| Intensity | Figure 13(b) | 0.49 | −0.38 | 0.0 |
| Line width | Figure 14(b) | 0.23 | −0.27 | 0.001 |

The double peaks in the Doppler velocity and centroid distribution could be due to the different nature of flows in different parts of the observed active region. In a recent study, Humphries et al. (2020) show redshift at the footpoints and blueshift near the top of the coronal protrusions. Studies by McIntosh & De Pontieu (2009) and De Pontieu et al. (2017) suggest that the outflows could be linked to chromospheric jets and spicules, which are heated while propagating into the corona. Solar wind flows are often considered to explain the active region outflows (Sakao et al. 2007). Brooks et al. (2011), Brooks & Warren (2012), and Brooks (2015) showed through EIS observations that the active region outflows contain slow wind plasma composition. The reported shifts may depend on various factors like the part of the active region they belong to, the phase of the solar cycle, the wavelength of the spectral line, the height of the corona, etc. Further studies are needed to ascertain the causes of the flow.

As mentioned above, the dependence of line width on Doppler velocity/centroid has been reported. Three out of four data sets presented in Section 3.3 have very good correlations. Raju et al. (2011) found that the correlation between Doppler velocity and line width is significant. Peter (2010) states that a positive correlation between Doppler velocity and line width indicates a heating process driving a flow. Line width broadening is found to be associated with strong blueshifts in general. The enhanced line broadening is reported to lead to the growth of Alfvén wave amplitude or inhomogeneity of flow velocities (McIntosh & De Pontieu 2009; Chen et al. 2010; Tian et al. 2012).

Lastly, the correlation of the intensity and the line width with the solar cycle is interesting and expected to some extent. Note that our data samples are not restricted to the disk or limb or the umbra or penumbra of the active regions, but are spread globally over the Sun. It is known that the EUV lines show an overall increase in intensity with the solar cycle. The increase in line width implies an increase of temperature and/or the nonthermal velocity with the solar cycle. This may have implications for the coronal heating.

## 5. Conclusions

Active regions from different parts of the Sun were analyzed for a period of 13 yr using Hinode/EIS data in the Fe XII $\lambda 195.12$ line. The physical parameters intensity, line width, Doppler velocity, and centroid are obtained from their line profiles. The behaviors of the parameters are studied through a multicomponent line profile simulation. We found that the Doppler velocity and centroid are not always related, but shift relatively in six different ways. First, the Doppler velocity and the centroid distributions were found to be centered around the rest wavelength (Case 1). Both were found to be shifted to the blue or red in Cases 2 and 3. Further, we found that the Doppler velocity was centered around the rest wavelength, while the centroid was found to shift to the blue or red partially or completely (Cases 4 and 5). Lastly, we found that the Doppler velocity was shifting to the red, while the centroid was in blue (Case 6). We found a majority of the data falling under Case 4 (46.1%), followed by Case 5 (26.62%), and then Case 1 (24.6%). The intensity distribution is skewed in most of the cases. The line width distribution is mostly symmetric but skewed in some cases. The average line width is found to be around 35 mÅ. The Doppler velocity and centroid are found to range between $-60$ and $60\,\mathrm{km\,s^{-1}}$ in most of the cases. In some extreme cases, the centroid is found to shift up to $-100$–$100\,\mathrm{km\,s^{-1}}$. The line width is found to follow a second-degree polynomial in its distribution with the Doppler velocity and the centroid. This is shown from the simulation as well. It is found that the line width increases with the increasing absolute value of Doppler velocity or centroid in most cases. Lastly, the intensity and the line width are found to show some dependence on the solar cycle, although there is no such correlation between the Doppler velocity and centroid with the solar cycle.

This work is funded by the Department of Science & Technology. M.P. would like to thank the Women Scientist Scheme—A (WOS-A) of the Department of Science & Technology for providing financial support to carry out this project.

This paper is based on data acquired from Hinode/EIS. Hinode is a Japanese mission developed and launched by ISAS/JAXA, collaborating with NAOJ as a domestic partner, NASA and STFC (UK) as international partners. Scientific operation of the Hinode mission is conducted by the Hinode science team organized at ISAS/JAXA. This team mainly consists of scientists from institutes in the partner countries. Support for the post-launch operation is provided by JAXA and NAOJ(Japan), STFC (UK), NASA, ESA, and NSC (Norway).

Disclosure of Potential Conflicts of Interest: The authors declare that they have no conflicts of interest.

## ORCID iDs

Maya Prabhakar  https://orcid.org/0000-0003-2141-4184
K. P. Raju  https://orcid.org/0000-0002-6648-4170

## References

Beckers, J. M. 1968, SoPh, 3, 367
Brekke, P., Thompson, W. T., Woods, T. N., & Eparvier, F. G. 2000, ApJ, 536, 959
Brooks, D. 2015, in AGU Fall Meeting (Washington, DC: AGU), SH54B-06
Brooks, D. H., Harra, L., Bale, S. D., et al. 2021, ApJ, 917, 25
Brooks, D. H., & Warren, H. P. 2012, ApJL, 760, L5
Brooks, D. H., Warren, H. P., & Young, P. R. 2011, ApJ, 730, 85
Chae, J., Yun, H. S., & Poland, A. I. 1998, ApJS, 114, 151
Chen, F., Ding, M. D., & Chen, P. F. 2010, ApJ, 720, 1254
Culhane, J. L., Harra, L. K., James, A. M., et al. 2007, SoPh, 243, 19
Dadashi, N., Teriaca, L., & Solanki, S. K. 2011, A&A, 534, A90
Delone, A. B., & Makarova, E. A. 1969, SoPh, 9, 116
Delone, A. B., & Makarova, E. A. 1975, SoPh, 45, 157
De Pontieu, B., Erdélyi, R., & James, S. P. 2004, Natur, 430, 536
De Pontieu, B., Hansteen, V. H., Rouppe van der Voort, L., van Noort, M., & Carlsson, M. 2007, ApJ, 655, 624
De Pontieu, B., Martinez-Sykora, J., De Moortel, I., Chintzoglou, G., & McIntosh, S. W. 2017, AGUFM, SH43A-2793
De Pontieu, B., McIntosh, S. W., Hansteen, V. H., & Schrijver, C. J. 2009, ApJL, 701, L1
Dere, K. P., Doschek, G. A., Mariska, J. T., et al. 2007, PASJ, 59, S721
Doschek, G. A., Warren, H. P., Mariska, J. T., et al. 2008, ApJ, 686, 1362
Hara, H., Watanabe, T., Harra, L. K., et al. 2008, ApJL, 678, L67
Harra, L., Brooks, D. H., Bale, S. D., et al. 2021, A&A, 650, A7
Hollweg, J. V. 1978, RvGeo, 16, 689
Humphries, L. D., Verwichte, E., Kuridze, D., & Morgan, H. 2020, ApJ, 898, 17
Klimchuk, J. A. 2006, SoPh, 234, 41
Klimchuk, J. A. 2012, JGRA, 117, A12102
Kosugi, T., Matsuzaki, K., Sakao, T., et al. 2007, SoPh, 243, 3
Macneil, A. R., Owen, C. J., Baker, D., et al. 2019, ApJ, 887, 146
Marsch, E., Tian, H., Sun, J., Curdt, W., & Wiegelmann, T. 2008, ApJ, 685, 1262
McIntosh, S. W., & De Pontieu, B. 2009, ApJL, 706, L80
McIntosh, S. W., De Pontieu, B., & Leamon, R. J. 2010, SoPh, 265, 5
McIntosh, S. W., Tian, H., Sechler, M., & De Pontieu, B. 2012, ApJ, 749, 60
Parker, E. N. 1988, ApJ, 330, 474
Parnell, C. E., & De Moortel, I. 2012, RSPTA, 370, 3217
Patsourakos, S., & Klimchuk, J. A. 2006, ApJ, 647, 1452
Peter, H. 2010, A&A, 521, A51
Peter, H., & Judge, P. G. 1999, ApJ, 522, 1148
Prabhakar, M., Raju, K. P., & Chandrasekhar, T. 2019, SoPh, 294, 26
Raju, K. P., Chandrasekhar, T., & Ashok, N. M. 2011, ApJ, 736, 164
Raju, K. P., Desai, J. N., Chandrasekhar, T., & Ashok, N. M. 1993, MNRAS, 263, 789
Sakao, T., Kano, R., Narukage, N., et al. 2007, Sci, 318, 1585
Taylor, J. R. 1982, An Introduction to Error Analysis. The Study of Uncertainties in Physical Measurements (Mill Valley, CA: Univ. Science Books)
Tian, H., Harra, L., Baker, D., Brooks, D. H., & Xia, L. 2021, SoPh, 296, 47
Tian, H., McIntosh, S. W., De Pontieu, B., et al. 2011, ApJ, 738, 18
Tian, H., McIntosh, S. W., Wang, T., et al. 2012, ApJ, 759, 144
Tian, H., Tu, C., Marsch, E., He, J., & Kamio, S. 2010, ApJL, 709, L88
Tripathi, D., & Klimchuk, J. A. 2013, ApJ, 779, 1
Yardley, S. L., Brooks, D. H., & Baker, D. 2021, A&A, 650, L10